\documentclass[twocolumn,showpacs,preprintnumbers,amsmath,amssymb]{revtex4} 
\usepackage{epsfig}
\usepackage{amsmath}
\usepackage{graphicx}
\usepackage{amssymb} 
\usepackage{graphics}
\usepackage{psfrag}
\usepackage{float}
\usepackage{latexsym}
\usepackage{exscale}
\usepackage{array}

\begin{document}
\title{Evidence of universality in the dynamical response of micromechanical diamond resonators at millikelvin temperatures}

\author{Matthias Imboden and Pritiraj Mohanty}
\affiliation{Department of Physics, Boston University, 590 Commonwealth Avenue, Boston, Massachusetts 02215, USA}
\email{mohanty@bu.edu}
\date{\today}

\begin{abstract}
We report kelvin to millikelvin-temperature measurements of dissipation and frequency shift in megahertz-range resonators fabricated from ultra-nanocrystalline diamond. Frequency shift $\delta f/f_0$ and dissipation $Q^{-1}$ demonstrate temperature dependence in the millikelvin range similar to that predicted by the glass model of tunneling two level systems.  The logarithmic temperature dependence of $\delta f/ f_0$ is in good agreement with such models, which include phonon relaxation and phonon resonant absorption. Dissipation shows a weak power law, $Q^{-1}\propto T^{\frac{1}{3}}$, followed by saturation at low temperature. A comparison of both the scaled frequency shift and dissipation in equivalent micromechanical structures made of single-crystal silicon and gallium arsenide indicates universality in the dynamical response.
\end{abstract}
\pacs{85.85.+j,62.25.-g,65.40.De,81.05.Uw}
\maketitle

\section{Introduction}
Micromechanical resonators are instrumental in the investigation of a wide variety of fundamental physics problems.  These include quantum measurement and quantum computation~\cite{bocko-onofrio,qnm}, ultra-sensitive force and mass detection~\cite{rugar-force}, single spin detection~\cite{rugar-spin}, gravitational wave detection~\cite{bocko-onofrio} and other fundamental phenomena~\cite{wu,montemagno}.  The relevant features of a resonator are characterized by the resonance frequency shift $\delta f$ (compared to the lowest temperature resonance frequency) and dissipation (inverse quality factor) $Q^{-1}$. Reduced dimensions are necessary for achieving high resonance frequencies. However, miniaturization beyond the sub-micron scale leads to a dramatic increase in the surface-to-volume ratio, resulting in increased dissipation, limiting device performance. To counter this trend it is therefore necessary to avoid extrinsic mechanisms and minimize intrinsic mechanisms. 
 
From the low temperature response of single crystal silicon resonators at both kilohertz~\cite{KleimanPRL1987} and megahertz~\cite{guitiPRB} frequencies, it is clear that low-lying energy excitations of internal defects or two-level systems (TLS) provide the dominant contribution to intrinsic dissipation in micromechanical resonators. Even though the general trend of the low temperature dependence is easily explained by the standard glass model of TLS~\cite{esquinaziSpringer1998, phillipsRPP1987}, additional experiments in other materials such as silicon~\cite{guitiPRB} and gallium arsenide~\cite{seungboAPL} and detailed theoretical calculations~\cite{cesar,ramos} suggest an incomplete understanding of the temperature dependence of the quality factor.

Here, we report a detailed set of low temperature measurements in a novel material, ultra-nano-crystalline diamond (UNCD). We find that the resonance frequency shift and quality factor of resonators fabricated from UNCD, single-crystal silicon, and epitaxially-grown gallium arsenide show universality in their temperature dependence. The universal behavior provides further evidence that intrinsic dissipation in micromechanical resonators is dominated by crystal defects, whether they are (substitutional) crystal impurities or (configurational) surface defects caused by the abrupt lattice termination. Our measurements indicate that the same TLS processes dominate in a range of materials, underlining the need for a single material independent theory.

Polycrystalline diamond is an exciting material for micromechanical devices due to its unique combination of physical and electrical properties.  The UNCD used here does not maintain the high thermal conductivity of single crystal and NCD (nanocrystalline diamond), and the measured Young's modulus is low compared to single crystal and NCD diamond materials.  In contrast, UNCD is distinguished by the ability to be grown to a wide range of thickness without compromising surface friction and roughness.  NCD the crystal size is generally on the same order as the film thickness and grows in size with film thickness.  UNCD maintains low surface roughness through a re-nuclearization  process that induces new crystals to form.  The resulting conglomeration of $5-10~nm$ sized crystals is extremely smooth up to considerable thicknesses.  In both NCD and UNCD electrical properties can be varied over a wide scale by controlling the doping concentration. Here, low-doped wafers are used, and the material is electrically insulating.

We find that dissipation follows a power law in temperature, $Q^{-1}\propto T^{\frac{1}{3}}$,  independent of resonance frequency and material.  At low temperatures, below $\approx 100~mK$, dissipation saturates.  Upon cooling, a shift in the sound velocity $\delta v/ v_0$ can be observed.  This is equivalent to measuring the frequency shift $\delta f/f_0$~\cite{topp}.  Below a characteristic temperature $T_\diamond\approx$ 1-2~K, where resonant-absorption processes of TLS are active, we observe a logarithmic temperature dependence with a positive slope.  Above $T_\diamond$, TLS relaxation-absorption mechanisms dominate. The temperature dependence is also logarithmic, however with a negative and steeper slope.  Interestingly, the slope of $\log(\delta f/f_0)$ scales with $f_0^{-1}$.  We find that data from sub-micron-sized mechanical resonators from UNCD, single crystal silicon and gallium arsenide heterostructures can be scaled onto a universal curve for both $\delta f / f_0$ and $Q^{-1}$.
\begin{figure}[htb]
\includegraphics[width=.5\textwidth]{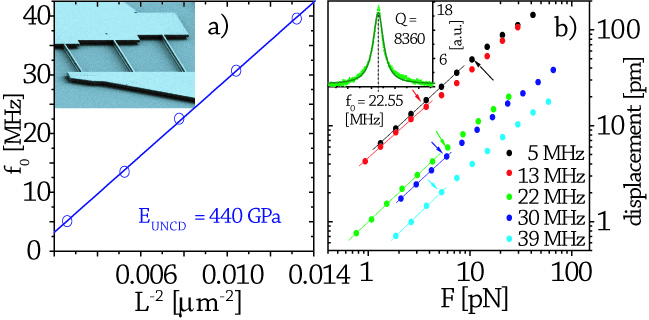}   
\caption{(color online) a) Frequency-length$^{-2}$ relation including a linear fit from which the Young's modulus of $440~\pm~70~GPa$ is determined.  Insert: SEM micrograph of the diamond harp structure with three doubly clamped beams. b) Displacement vs. force plot for all five fundamental resonances, the spring constant is given by the inverse of the slope. The arrows mark the onset of non-linearity. Insert: 22.55~MHz resonance with lorentzian fit of the 11.3~$\mu m$ beam.}
\label{figure1}
\end{figure}

\section{Experimental Setup}
The micromechanical resonators (Fig.~\ref{figure1} a) are fabricated using standard micro-machining procedures (e-beam lithography, metalization, reactive ion etching and finally an HF wet-etch to release the structures).  The samples are etched from 340~nm thick Aqua25 films obtained from Advanced Diamond Technologies, Inc.  Mechanical resonances from five doubly-clamped beam structures of 350~nm width and lengths varying from 8.7-19.6~$\mu$m are studied.  The structures are actuated and detected by the magnetomotive technique~\cite{mohantyPRB2002} in high vacuum.  Resonance frequencies $f_0$ and dissipation $Q^{-1}=\Delta f/f_0$ are obtained from a lorentzian fit (illustrated for the 22.55~MHz resonance in Fig.~\ref{figure1} b), $\Delta f$ is the full width at half maximum).  

The length dependence of frequency is depicted for the five structures in Fig.~\ref{figure1} c).  Inclusion of the metallic electrodes on top of the diamond modifies the resonance frequency:  $f_0=\eta\sqrt{(E_d I_d+E_m I_m)/(\rho_d A_d+\rho_m A_m)}L^{-2}$~\cite{yangAPL2001}, where $E_i, I_i, \rho_i$, and $A_i$ are the Young's modulus, the second moment of inertia, the density and the cross sectional area respectively of both the diamond and metal.  $L$ is the beam length and $\eta$ a numerical factor.  The precise fit validates the use of the thin-beam approximation. From the slope, we determine the Young's modulus of the thin film to be 440~GPa. This is considerably lower than expected, when compared to the Young's modulus previously measured in polycrystalline diamond~\cite{imbodenAPL2007}.  It is common that thin film diamond contains a large number of $sp^2$ bonds on grain boundaries, which degrade the Young's modulus.  UNCD is particularly sensitive to this as the high density of nano-crystal grains effectively increase the area of the surface where the $sp^2$ bonds form. Compressive strain within the film may arise with cooling do to the thermal expansion mismatch of polycrystalline diamond and the silicon base. The frequency shift caused by tensions is discussed by Postma et al.~\cite{postma}. Figure~\ref{figure1} d) shows the oscillation amplitude on resonance, $x(f_0)=Q F_{dr}/k_{eff}$ as a function of the drive force, $F_{dr}=BLI_{dr}$ is the actuation voltage) for all five beam resonances, where the measured magnetomotive response, $V_{emf}(\omega_0)=\frac{\xi L^2 B^2 Q}{m\omega_0}I_{dr}(\omega_0)$, is proportional to the beam amplitude~\cite{mohantyPRB2002}, where $B$ is the applied perpendicular magnetic field, $m$ the effective mass of the resonator, $\omega_0$ the resonance frequency and $I_{dr}$ the drive current across the beam.  $\xi$ is an integration factor.  The effective spring constant $k_{eff}$ is determined from the inverse of the slope of the plot in figure~\ref{figure1} b) and found to be on order unity for all structures. For the following data, care is taken to drive the resonators only in the linear regime.

\begin{figure*}[htb]
\includegraphics[width=0.9\textwidth]{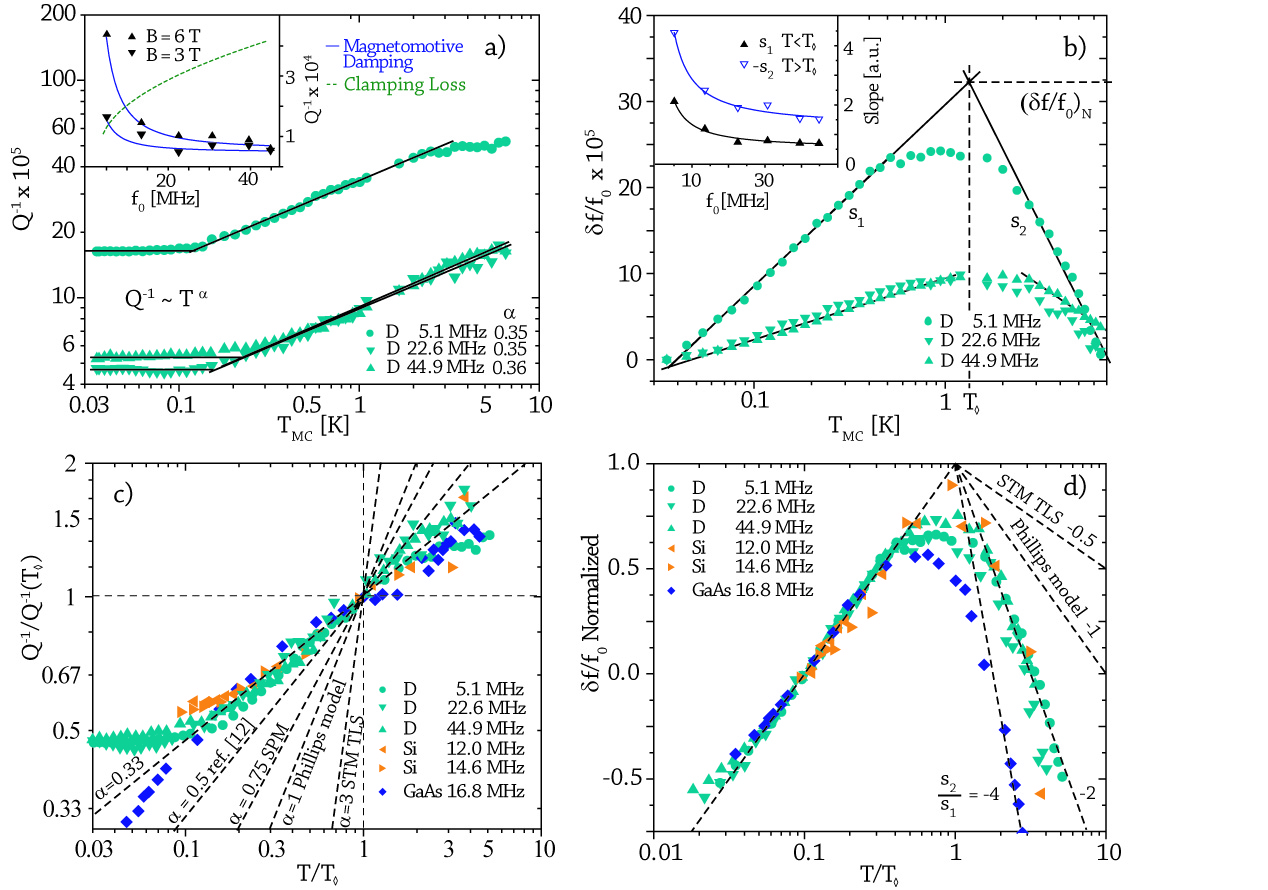} 
\caption{(color online)  a) $Q^{-1} - T$ dependency for three resonance frequencies at 3~T.  We measure $Q^{-1} \propto T^{\alpha}$, where $\alpha \approx \frac{1}{3}$.  Below 100 mK dissipation saturates.  Insert: Dissipation-frequency relation at 6~T and 3~T, including fits for magnetomotive damping ($Q_{MMD}^{-1}\propto f_0^{-\frac{3}{2}}$) and clamping loss ($Q_{CL}^{-1}\propto f_0^{\frac{1}{2}}$).   $T_{MC}$ is the mixing chamber temperature.  b)  $\delta f / f_0 - T$ dependency, $f_0$ is the resonance frequency at 35~mK.  Below and above a characteristic temperature $T_\diamond$ the shift is logarithmic as predicted by TLS models. The insert depicts the slopes of $\delta f / f_0$ and $f_0^{-1}$ fit for low (up triangles) and high (down triangles) temperature regimes. c) Normalized dissipation $Q^{-1}/Q^{-1}_{T_\diamond}$ for UNCD, Si~\cite{guitiPRB}, and GaAs~\cite{seungboAPL} for frequencies ranging from 5~MHz to 45~MHz.  The temperature has been scaled to $T_\diamond$ (see panel b)).  The black dotted line represents a slope of $\frac{1}{3}$, alluding to the same power law for all three materials. The other slopes represent predictions of various TLS models.  d) $(\delta f / f_0)_N$ vs. $T/T_\diamond$. $f_0$ is the frequency measured at $0.1~T_\diamond$. The curves collapse if the ratio $s_2/s_1$ is the same. The dashed lines illustrate four $\frac{s_2}{s_1}$ ratios.  $\frac{s_2}{s_1}=-0.5$ and $\frac{s_2}{s_1}=-1$ depicts the predictions of the TLS standard tunneling model~\cite{esquinaziSpringer1998} and the Phillips model~\cite{ phillipsRPP1987} respectively.}
\label{figure2and3}
\end{figure*}
\section{Results}
There are a number of dissipation mechanisms that are known to contribute to loss in micromechanical resonators.  Figure~\ref{figure2and3} a) depicts the temperature dependence for three resonance frequencies; the insert illustrates how dissipation varies with frequency at two applied magnetic field strengths.  Short, doubly-clamped beams often suffer from clamping losses and dissipation is dominated by energy loss into the pads~\cite{imbodenAPL2007}.  For this case, dissipation rapidly grows with decreasing resonator length.  We observe the opposite behavior, and hence, we can rule out clamping loss as a dominant dissipation mechanism.  Thermoelastic damping (self-heating) decreases with frequency; this mechanism however has been shown not to contribute at low temperatures and small structure sizes~\cite{roszhart1990IEEE}.  Internal losses in the electrode also contribute to dissipation:  $Q^{-1}=\left[1/(1+\beta)\right](Q_{d}^{-1}+\beta Q_{Au}^{-1})$ with $\beta = (E_{Au} t_{Au})/(E_{d} t_{d})=0.02$, where $t_i$ is the film thickness of the gold and diamond. A conservative estimate for losses in the gold electrode predicts an effect smaller than $5\%$.  It can be shown that magnetomotive damping (a dissipation mechanism caused by the interaction of the readout curcuit and an externally applied magnetic field)~\cite{cleland} is proportional to $f_0^{-1.5}$.  The insert of Fig.~\ref{figure2and3} a) shows good agreement with this prediction for applied fields of 6~T and 3~T. At high frequencies, magnetomotive damping no longer contributes significantly as indicated by the saturation of the data.  For the following dissipation data discussed in this letter, dissipation measurements are taken at 3~T, in the magnetic field regime where magnetomotive damping can be neglected. For the frequency shift data, measurements are taken at 6~T, as no significant field dependence of this variable is observed.  

The presence of TLSs become apparent through the temperature dependence of both dissipation, and, in particular, resonance frequency, displayed in Fig.~\ref{figure2and3} a) and b) respectively. Dissipation saturates below $100~mK$ with a possible recurrence beginning at temperatures above 2-4~K, observable for the lowest frequency.  Within this temperature range, dissipation follows a power law,  $Q^{-1}\propto T^{\alpha}$ with $\alpha \approx 0.35$, and no observed frequency correlation.  At higher field strengths, where magnetomotive damping and possibly magnetic field-TLS coupling occurs, the temperature dependence becomes weaker, while the decreasing trend is strongest for lower frequencies (data not shown).  The lack of a clear frequency dependence of $\alpha$ measured at 3~T indicates that the measurements were taken in the parameter space where field coupling can be neglected.

The frequency shifts vs. temperature plots in Fig.~\ref{figure2and3} b) show a logarithmic dependence, where at $T_\diamond$ the slope changes sign and increases in magnitude.  The logarithmic behavior is in good agreement with glassy TLS models, however the slope is expected to remain constant in magnitude above $T_\diamond$ (Phillips model~\cite{phillipsRPP1987}).  The fit results are summarized in Table~\ref{table1}. We observe that the magnitude of the slope $s_i$ doubles for $T>T_\diamond$ independent of frequency, where $s_1$ is the slope for $T<T_\diamond$ and $s_2$ is the slope for $T>T_\diamond$.  The slopes show an inverse dependence on frequency, also not predicted by standard theories. This behavior has been observed by Kleiman et al.~\cite{KleimanPRL1987} in previous work on single crystal silicon resonators as well.  $T_\diamond$ shows a rising trend with increasing resonance frequency. A precise dependence could not be determined from the available data.
\begin{table}[htb]\footnotesize
\caption{Fit results for $\delta f/f_0 - T$ dependency for both temperature regimes ($T<T_\diamond \ s_1$, $T>T_\diamond \ s_2$). The diamond and silicon resonators show roughly similar slope ratios $-s_2/s_1\approx 2$, while for gallium arsenide the magnitude of this ratio significantly greater $-s_2/s_1\approx 4$.  The $s_2$ values quoted for silicon are rough estimates. $T_\diamond$ is defined in Fig.~\ref{figure2and3} b). Errors on the frequency are on the order of $10~Hz$. The error in $T_\diamond$ is considerable due to the limited data points, especially above $T_\diamond$.  This uncertainty is estimated to be $\pm 0.25~K$, hence the scatter is explained by experimental error.}
\begin{tabular*}{0.48\textwidth}{@{\extracolsep{\fill}}cccccc}
\hline\hline
Material & $f_0$   & $s_1$              & $s_2$              & $-\frac{s_2}{s_1}$    & $T_\diamond$       \\
     &$[$MHz$]$& $\times10^{-5}$    & $\times10^{-5}$        &                       & [K]                \\ \hline  
UNCD &5.11     &  21.2              & -44.3                  &  2.09                 & 1.27   		        \\
UNCD &13.50    &  12.0              & -24.8                  &  2.07                 & 1.86    					  \\
UNCD &22.55    &  7.5               & -19.0              		 &  2.53                 & 1.61   						\\
UNCD &30.71    &  8.0               & -19.9                  &  2.48                 & 1.91    						\\
UNCD &39.59    &  7.2               & -15.3                  &  2.12                 & 1.71   					  \\
UNCD &44.85    &  7.1               & -15.1                  &  2.11                 & 1.92  					    \\ \hline
Si   &12.03    &	1.8   						& (-4.6)                 & (2.59)                & 0.54               \\
Si   &14.59 	 &	2.0   						&	(-3.1)                 & (1.59)            	   & 0.64               \\ \hline
GaAs &15.82	   &	13.5	  					&	-55.7                  &	4.13				   		   & 1.29               \\
\hline\hline
\end{tabular*}
\label{table1}
\end{table}
\section{Discussion}
At low temperatures, TLS are confined to a nearly degenerate ground state, modeled by an asymmetric double well potential with asymmetry $\Delta$ and tunnel splitting energies $\Delta_0$, typically on the order of 1~K. Glass models generally assume a constant density of states, independent of $\Delta$ and $\Delta_0$, given by $P(\Delta,\Delta_0)d\Delta d\Delta_0=\frac{P_0}{\Delta_0}d\Delta d\Delta_0$.  This model gives an expression to determine the number of defects from the coefficient of the logarithmic dependency of the frequency shift with temperature. however, it is assumed that this slope is frequency (length) independent, and hence does not apply. It may still be useful to use this as a first approximation, where one obtains on the order of 400 defects for the shortest to 3200 defects for the longest beam resonators. The Phillips TLS model is a more constrained model with a fixed splitting energy and a Gaussian asymmetry distribution with width $\Delta_1$, while the density of states takes the form $P(\Delta)d\Delta= A\exp (-\Delta^2/2\Delta_1^2)d\Delta$.  For both models the logarithmic dependence of the frequency shift is reproduced, with different coefficients.  The Soft potential Model (SPM) assumes a quartic potential and constant density of states for each mode, based on the concept of localized low frequency sound waves. The SPM is successful in explaining low temperature thermal conductivity and predicts a $Q^{-1} \propto T^\frac{3}{4}$ dissipation law with low temperature saturation~\cite{ramos}. Standard single crystal models predict non-logarithmic temperature dependence of the frequency shift and non-power-law temperature dependence of the dissipation, in contrast to the data presented here. These theoretical results are summarized in~\cite{guitiPRB}. Considering the large surface to volume ratio of UNCD one expects a high density of TLSs.  Theories have been developed to take interacting TLSs into account~\cite{esquinaziSpringer1998},  these too however fail to explain the observed behavior.

The temperature dependence for both frequency shift and dissipation scales to a universal curve for all the three materials studied in this temperature range. For scaling, the measured temperature is referenced to the characteristic temperature $T_\diamond$ defined in Fig.~\ref{figure2and3} b).  All dissipation plots are normalized to the dissipation value measured at $T_\diamond$, $Q^{-1}/Q^{-1}_{T_\diamond}$ and plotted vs. $T/T_\diamond$.  Figure~\ref{figure2and3} c) illustrates the power law $Q^{-1}\propto T^{\frac{1}{3}}$ for the three materials. Low temperature saturation is observed in diamond as well as in silicon, however this is not seen in the gallium arsenide resonators for the lowest temperatures measured. Correspondingly the $\frac{s_2}{s_1}\approx -4$ for gallium arsenide, is significantly greater from that measured in diamond and silicon. To compare the shift in frequency for different temperatures, the data are normalized to $(\delta f/f_0)_N$, defined in Fig.~\ref{figure2and3} b) and plotted vs. $T/T_\diamond$.  For these plots $f_0$ is defined as the frequency at $0.1~T_\diamond$.  Although there are not enough temperature data points for the silicon resonators to make statements with high confidence, it appears that the three materials behave qualitatively the same, as above $T_\diamond$ the slope of $\delta f/f_0$ not only changes sign but also increases in magnitude (see Fig.~\ref{figure2and3} d)). Such observations are consistent with data in literature for a wide range of structures and frequencies~\cite{KleimanPRL1987,esquinaziSpringer1998,phillipsRPP1987}.  

The large surface-to-volume ratio introduces a significant amount of defects, concealing the presence of a crystal; hence the micromechanical resonators behave like glassy structures.  The data sets appear to fall into universality classes defined by the $s_2/s_1$ ratio; this must still be confirmed with future experiments. The universal behavior of these micromechanical devices motivates the need for a theory that captures more accurately the observed relations.  A rigorous theoretical framework must be developed that encompasses the density of states of TLSs, surface effects due to the high surface to volume ratio of micro-electromechanical resonators, and possibly a non-ohmic thermal bath. It remains to be seen if, with a variation of current TLS models, it is possible to theoretically reproduce simultaneously both the dissipation power law and the logarithmic frequency shift.  

While dissipation has been studied extensively at lower frequencies in single crystal materials such as silicon and gallium arsenide, we present first measurements down to millikelvin temperatures of micromechanical resonators fabricated from a nanocrystalline material.  Understanding phonon-dynamics (phonon coupling, role of disorder, crystal structure) in diamond is fundamentally important to novel condensed matter phenomena such as superconductivity in boron doped diamond and the pursuit of macroscopic quantum states in mechanical oscillators~\cite{bocko-onofrio,lahaye}.  A necessary first step is to understand phonon dynamics in undoped diamond.  What is fundamentally exciting is that such micro-resonators demonstrate the same universal glassy behavior as those made of silicon and gallium arsenide.

In conclusion, we report the presence of TLSs in UNCD diamond resonators at megahertz frequencies. Both the resonant absorption and relaxation absorption regimes are observed with a characteristic temperature around 1-2~K.  A constant power law in dissipation and analogous logarithmic frequency dependence on temperature is found and compared to measurements in silicon and gallium arsenide micromechanical resonators.  Further theoretical work is required to fully explain this universal behavior, possibly by taking more involved TLS density of states into account.

We thank Guiti Zolfagharkhani et al. and Seung Bo Shim et al. for making their data available to us.  We thank Oliver Williams, Etienne Bustarret and Eduardo Lopez for helpful discussions. This work is supported by the National Science Foundation (DMR-0449670).

\end{document}